\documentclass[review]{elsarticle}

\usepackage{hyperref}
\usepackage{graphicx,color,dcolumn,booktabs,bm}
\journal{Journal of \LaTeX\ Templates}

\usepackage{amsfonts}
\usepackage{amsmath}
\usepackage{graphicx}
\usepackage{subfigure}
\usepackage{dcolumn}
\usepackage{bm}
\usepackage{booktabs}
\usepackage[utf8]{inputenc}

\usepackage{graphicx,graphics,dcolumn,booktabs,bm}
\usepackage{longtable,lscape}
\usepackage{txfonts}
\usepackage{overpic}
\usepackage{amssymb}
\usepackage{indentfirst}
\usepackage{epsfig}
\usepackage{feynmf}   
\usepackage{epstopdf}   
\usepackage{slashed}  
\usepackage{multirow}

\def\GGa{\langle GG\rangle}
\def\GGb{\langle g_s^2GG\rangle}

\def\f(s){[(\alpha+\beta)m^2-\alpha\beta s]}
\def\non{\\ \nonumber}

\def\f(Q2){\left[\frac{m_Q^2(1-z)}{1-y}+\frac{m_Q^2(1-z)}{(1-x)xy}+m_Q^2z-Q^2z(1-z)\right]}

\begin{document}

\begin{frontmatter}

\title{Hunting for exotic doubly hidden-charm/bottom tetraquark states}

\author[Saskatoon]{Wei Chen}
\author[beihang]{Hua-Xing Chen}
\ead{hxchen@buaa.edu.cn}
\author[lanzhou1,lanzhou2]{Xiang Liu}
\ead{xiangliu@lzu.edu.cn}
\author[Saskatoon]{T. G. Steele}
\ead{tom.steele@usask.ca}
\author[pku1,pku2,pku3]{Shi-Lin Zhu}
\ead{zhusl@pku.edu.cn}

\address[Saskatoon]{Department of Physics and Engineering Physics, University of Saskatchewan, Saskatoon, Saskatchewan, S7N 5E2, Canada}
\address[beihang]{School of Physics and Beijing Key Laboratory of Advanced Nuclear Materials and Physics, Beihang University, Beijing 100191, China}
\address[lanzhou1]{School of Physical Science and Technology, Lanzhou University, Lanzhou 730000, China}
\address[lanzhou2]{Research Center for Hadron and CSR Physics, Lanzhou University and Institute of Modern Physics of CAS, Lanzhou 730000, China}
\address[pku1]{School of Physics and State Key Laboratory of Nuclear Physics and Technology, Peking University, Beijing 100871, China}
\address[pku2]{Collaborative Innovation Center of Quantum Matter, Beijing 100871, China}
\address[pku3]{Center of High Energy Physics, Peking University, Beijing 100871, China}

\begin{abstract}
We develop a moment QCD sum rule method augmented by fundamental inequalities to study the existence
of exotic doubly hidden-charm/bottom
tetraquark states made of four heavy quarks. Using the compact diquark-antidiquark configuration, we calculate
the mass spectra of these tetraquark states.
There are 18 hidden-charm $cc\bar c\bar c$ tetraquark currents with $J^{PC} = 0^{++}$, $0^{-+}$, $0^{--}$, $1^{++}$, $1^{+-}$, $1^{-+}$, $1^{--}$, and $2^{++}$.
We use them to perform QCD sum rule analyses, and the obtained masses are all higher than the spontaneous dissociation thresholds of two charmonium mesons,
which are thus their dominant decay modes.
The masses of the corresponding hidden-bottom $bb\bar b\bar b$ tetraquarks are all below or very close to the thresholds of
the $\Upsilon(1S)\Upsilon(1S)$ and $\eta_b(1S)\eta_b(1S)$, except one current of $J^{PC}=0^{++}$.
Hence, we suggest to search for the doubly hidden-charm states in the $J/\psi J/\psi$ and $\eta_c(1S)\eta_c(1S)$ channels.
\end{abstract}


\end{frontmatter}

In QED there exist the multi-lepton bound states composed of $e^+e^-e^+e^-$ and
$e^+e^-\mu^+\mu^-$ \cite{1947-Hylleraas-p493-496,2007-Cassidy-p195-197}. While, in QCD
the scalar mesons below 1 GeV may have the flavor configurations $q\bar q q \bar q$ and $q\bar q s\bar s$
\cite{1977-Jaffe-p281-281,2007-Chen-p94025-94025,Black:1998wt},
so are possible multiquark states. Actually,
at the birth of the quark model, these multiquark states beyond conventional mesons and baryons were proposed
by Gell-Mann \cite{1964-Gell-Mann-p214-215} and Zweig \cite{1964-Zweig-p-}. Searching for multiquark
matter has been an extremely intriguing research issue in the past fifty years, since it provides important
hints to deepen our understanding of the non-perturbative QCD~\cite{Chen:2016qju}. With experimental progress, more
and more candidates of multiquark matter were reported over the past decade, including
dozens of charmonium-like $XYZ$ states~\cite{2014-Olive-p90001-90001}, the hidden-charm pentaquarks
$P_c(4380)$ and $P_c(4450)$ \cite{2015-Aaij-p72001-72001}, and the $X(5568)$ observed by the D0 Collaboration~\cite{D0:2016mwd} but not confirmed in the LHCb~\cite{2016-Aaij-p152003-152003} and
CMS~\cite{2016-Collaboration-p-a} experiments. Hence, we conclude now to be the suitable time to hunt for
more exotic multiquark states.

When studying multiquark matter, distinguishing these possible multiquark states from the conventional
hadrons is a crucial task to establish their exotic nature.
An explicit road map to achieve this aim
is to study those exotic multiquark states with masses far away from the mass range of the observed conventional hadrons.
We note that the $J/\psi$ pairs were observed in LHCb~\cite{2012-Aaij-p52-59}, D\O~\cite{2014-Abazov-p111101-111101}
and CMS~\cite{2014-Khachatryan-p94-94}; the simultaneous $J/\psi\Upsilon(1S)$ events were recently observed in
D\O~\cite{2016-Abazov-p82002-82002} and CMS~\cite{CMS1}; the simultaneous $\Upsilon(1S)\Upsilon(1S)$ events
were also recently observed in CMS~\cite{2017-Khachatryan-p13-13}. All these events may be used to investigate the
four-heavy-quark ($QQ\bar Q \bar Q$) states, which fully satisfy this condition, and so are very good multiquark candidates.

In this letter we develop a moment QCD sum rule method augmented by fundamental inequalities to systematically study the doubly hidden-charm/bottom tetraquark states $cc\bar c\bar c$ and $bb\bar{b}\bar{b}$ with quantum numbers $J^{PC} = 0^{++}$, $0^{-+}$, $0^{--}$, $1^{++}$, $1^{+-}$, $1^{-+}$, $1^{--}$, and $2^{++}$.
In our study, we find a typical common peculiarity of the predicted masses of these doubly
hidden-charm and hidden-bottom tetraquark states, i.e., the masses of all $cc\bar c\bar c$ tetraquarks
are higher than the two charmonium thresholds while the $bb\bar{b}\bar{b}$ tetraquark states (except one
state of $J^{PC}=0^{++}$) lie below the dissociation thresholds of two bottomonium states, and these masses do not overlap with the mass ranges of the observed charmonia/bottomonia.
In addition to perform the mass spectrum analysis, their OZI-allowed strong decay behaviors are also predicted in the present work,
which are useful for future experimental searching.

These doubly hidden-charm/bottom $QQ\bar Q\bar Q$ tetraquark states did not receive much attention because of the absence of experimental
data, unlike the (singly) hidden-charm/bottom
$qQ\bar q\bar Q$ tetraquark system, where $q$ represents the light quark and $Q$ the heavy quark
\cite{2010-Chen-p105018-105018,2011-Chen-p34010-34010,2015-Chen-p54002-54002}.
However, with the running of LHC at 13 TeV and the forthcoming BelleII,
searching for these doubly hidden-charm and hidden-bottom tetraquark states
will probably become the next potential experimental issue in the near future.
Systematic theoretical studies of the doubly hidden-charm/bottom tetraquark states
seem to be imperative and useful for future experimental studies. However, the light
mesons ($\pi$, $\rho$, $\omega$, $\sigma$, $\cdots$) can not be exchanged between the two charmonium/bottomonium states,
which limits the use of many phenomenological methods~\cite{Chen:2016qju}.
To study these doubly hidden-charm/bottom tetraquark states, our present calculations in QCD sum rules and a very recent work
in the framework of the color-magnetic interaction \cite{Wu:2016vtq} are both crucial to shed light on
the long-standing debate on whether there exist such tetraquark states \cite{1975-Iwasaki-p492-492,1981-Chao-p317-317,1982-Ader-p2370-2370,1983-Ballot-p449-451,
1985-Heller-p755-755,2004-Lloyd-p14009-14009,1992-Silvestre-Brac-p2179-2189,1993-Silvestre-Brac-p273-282,
2006-Barnea-p54004-54004,2012-Berezhnoy-p34004-34004}.
We note that the traditional SVZ QCD sum rules \cite{1979-Shifman-p385-447,1985-Reinders-p1-1,2000-Colangelo-p1495-1576},
which we used to study the (singly) hidden charm pentaquarks $P_c(4380)$ and $P_c(4450)$ \cite{2015-Chen-p172001-172001},
can not be easily applied to study these doubly hidden-charm/bottom states, because of the four heavy quarks.
It is another version of QCD sum rules, the moment QCD sum rules \cite{1979-Shifman-p385-447,1985-Reinders-p1-1,2000-Colangelo-p1495-1576}, which we find to be capable of dealing with these states.
The moment sum rules have been very successfully used for reproducing the charmonium and bottomonium mass
spectra at the point $Q^2_0=0$ \cite{1979-Shifman-p385-447,1979-Shifman-p448-518,1983-Nikolaev-p526-526} and
$Q^2_0>0$ \cite{1981-Reinders-p109-109,1985-Reinders-p1-1}
and determining the strong coupling constant $\alpha_s$ and the heavy quark masses \cite{1997-Jamin-p334-352,2001-Eidemuller-p203-210,2001-Kuhn-p588-602}. Therefore, we shall use the moment QCD sum rules method in this letter
to systematically study the doubly hidden-charm/bottom states as the first theoretical calculation for such systems in QCD.

To start our study, we first construct local (no derivative operators) tetraquark interpolating currents with four heavy quarks $cc\bar c\bar c$ and $bb\bar b\bar b$ in the diquark-antidiquark configuration. Consisting of the same heavy quarks, the flavor structures for both the diquark $QQ$ and anti-diquark $\bar Q\bar Q$ are symmetric. The color structures are then fixed by the Fermi statistics for various Lorentz bilinear operators, i.e., symmetric $\mathbf{6_c}$ for the diquark fields $Q^T_a CQ_b (0^-)$, $Q^T_a C\gamma_5Q_b (0^+)$, $Q^T_aC\gamma_\mu\gamma_5Q_b (1^-)$,
and antisymmetric $\mathbf{\bar 3_c}$
for $Q^T_aC\gamma_\mu Q_b (1^+)$, $Q^T_a C\sigma_{\mu\nu}Q_b$ and $Q^T_aC\sigma_{\mu\nu}\gamma_5Q_b$, where $a, b$ are color indices.
Following Refs.~\cite{2013-Du-p14003-14003,2013-Chen-p2628-2628}, we obtain the color singlet tetraquark operators with four heavy quark fields and definite $J^{PC}$ quantum numbers outlined in the
following, where the colour structure can
be either symmetric $\mathbf{6_c}\times \mathbf{\bar 6_c}$ or antisymmetric $\mathbf{\bar 3_c}\times \mathbf{3_c}$.

The interpolating currents with $J^{PC}=0^{++}$ are
\begin{equation}
\begin{split}
J_1&=Q^T_aC\gamma_5Q_b\bar{Q}_a\gamma_5C\bar{Q}_b^T\, , \\
J_2&=Q^T_aC\gamma_\mu\gamma_5Q_b\bar{Q}_a\gamma^\mu \gamma_5C\bar{Q}_b^T\, , \\
J_3&=Q^T_aC\sigma_{\mu\nu}Q_b\bar{Q}_a\sigma^{\mu\nu}C\bar{Q}^T_b\, , \label{currents2} \\
J_4&=Q^T_aC\gamma_\mu Q_b\bar{Q}_a\gamma^\mu C\bar{Q}_b^T\, , \\
J_5&=Q^T_aCQ_b\bar{Q}_aC\bar{Q}_b^T\, .
\end{split}
\end{equation}
The interpolating currents with $J^{PC}=0^{-+}$ and $0^{--}$ are
\begin{equation}
\begin{split}
J_1^{\pm}&=Q^T_aCQ_b\bar{Q}_a\gamma_5C\bar{Q}_b^T\pm Q^T_aC\gamma_5Q_b\bar{Q}_aC\bar{Q}_b^T\, , \\
J_2^+&=Q^T_aC\sigma_{\mu\nu}Q_b\bar{Q}_a\sigma^{\mu\nu}\gamma_5C\bar{Q}^T_b\, , \label{currents1}
\end{split}
\end{equation}
in which $J^+_1$ and $J_2^+$ couple to the states with $J^{PC}=0^{-+}$, and $J_1^-$ couples to the states with $J^{PC}=0^{--}$.
The interpolating currents with $J^{PC}=1^{++}$ and $1^{+-}$ are
\begin{align}
J_{1\mu}^{\pm}&=Q^T_aC\gamma_\mu\gamma_5 Q_b\bar{Q}_aC\bar{Q}_b^T
\pm Q^T_aCQ_b\bar{Q}_a\gamma_\mu\gamma_5 C\bar{Q}_b^T\, , \label{currents4}
\non
J_{2\mu}^{\pm}&=Q^T_aC\sigma_{\mu\nu}\gamma_5 Q_b\bar{Q}_a\gamma^\nu C\bar{Q}^T_b
\pm Q^T_aC\gamma^\nu Q_b\bar{Q}_a\sigma_{\mu\nu}\gamma_5C\bar{Q}^T_b\, ,
\end{align}
in which $J_{1\mu}^{+}$ and $J_{2\mu}^{+}$ couple to the states with $J^{PC}=1^{++}$, and $J_{1\mu}^{-}$ and $J_{2\mu}^{-}$ couple to the states with $J^{PC}=1^{+-}$.
The interpolating currents with $J^{PC}=1^{-+}$ and $1^{--}$ are
\begin{equation}
\begin{split}
J_{1\mu}^{\pm}&=Q^T_aC\gamma_\mu\gamma_5 Q_b\bar{Q}_a\gamma_5C\bar{Q}_b^T
\pm Q^T_aC\gamma_5Q_b\bar{Q}_a\gamma_\mu\gamma_5 C\bar{Q}_b^T \, ,\\
J_{2\mu}^{\pm}&=Q^T_aC\sigma_{\mu\nu}Q_b\bar{Q}_a\gamma^\nu C\bar{Q}^T_b
\pm Q^T_aC\gamma^\nu Q_b\bar{Q}_a\sigma_{\mu\nu}C\bar{Q}^T_b \, ,\label{currents3}
\end{split}
\end{equation}
in which $J_{1\mu}^{+}$ and $J_{2\mu}^{+}$ couple to the states with $J^{PC}=1^{-+}$, and $J_{1\mu}^{-}$ and $J_{2\mu}^{-}$ couple to the states with $J^{PC}=1^{--}$.
The interpolating current with $J^{PC}=2^{++}$ is
\begin{align}
J_{1\mu\nu}&=Q^T_aC\gamma_\mu Q_b\bar{Q}_a\gamma_\nu C\bar{Q}_b^T
+Q^T_aC\gamma_\nu Q_b\bar{Q}_a\gamma_\mu C\bar{Q}_b^T\, , \label{currents5}
\non
J_{2\mu\nu}&=Q^T_aC\gamma_\mu\gamma_5 Q_b\bar{Q}_a\gamma_\nu\gamma_5 C\bar{Q}_b^T
+Q^T_aC\gamma_\nu\gamma_5Q_b\bar{Q}_a\gamma_\mu\gamma_5 C\bar{Q}_b^T\, .
\end{align}

In the following, we will explore the doubly hidden-charm/bottom tetraquark
systems in the framework of moment QCD sum rules.
We start with the two-point correlation functions
\begin{equation}
\begin{split}
\Pi(q)&= i \int d^4xe^{iq\cdot x}\langle 0|T[J(x)J^{\dag}(0)]|0\rangle\, , \\
\Pi_{\mu\nu}(q)&=i\int d^4x e^{iq\cdot x}\langle 0|T [J_\mu(x)J_\nu^\dagger(0)]|0\rangle\, , \\
\Pi_{\mu\nu, \,\rho\sigma}(q)&=i\int d^4x e^{iq\cdot x}\langle 0|T [J_{\mu\nu}(x)J_{\rho\sigma}^\dagger(0)]|0\rangle\, ,
\end{split}
\end{equation}
in which
the interpolating currents $J(x)$, $J_{\mu}(x)$ and $J_{\mu\nu}(x)$ couple to the scalar, vector and tensor states via
the following relations respectively:
\begin{equation}
\begin{split}
\langle0|J|X\rangle&=f_X\, ,
\\
\langle0|J_{\mu}|Y\rangle&=f_Y\epsilon_{\mu}\, ,
\\
\langle0|J_{\mu\nu}|Z\rangle&=f_Z\epsilon_{\mu\nu}\, ,  \label{coupling parameters}
\end{split}
\end{equation}
where $f_X$, $f_Y$ and $f_Z$ are coupling constants between interpolating currents and hadron states.
$\epsilon_{\mu}$ and $\epsilon_{\mu\nu}$ are the polarization vector and tensor, respectively.

The two-point correlation function can be expressed in the form of the dispersion relation at the hadron level
\begin{align}
\Pi(q^2)=\frac{(q^2)^N}{\pi}\int_{M_H^2}^{\infty}\frac{\mbox{Im}\Pi(s)}{s^N(s-q^2-i\epsilon)}ds+\sum_{n=0}^{N-1}b_n(q^2)^n\, ,
\label{dispersionrelation}
\end{align}
where $M_H$ is the mass of the physical state and $b_n$ are unknown subtraction constants. The imaginary part of the correlation function can be written
as a sum over $\delta$ functions,
\begin{align}
\nonumber
\text{Im}\Pi(s)&=\pi\sum_n\delta(s-m_n^2)\langle0|J|n\rangle\langle
n|J^{\dagger}|0\rangle+\text{continuum} \\
&=\pi f_X^2\delta(s-m_X^2)+\text{higher states}+\text{continuum}\,  \label{Imaginary}
\end{align}
in which a narrow resonance approximation is adopted since the lowest resonance is usually very sharp.
The spectral function can thus be parameterized as a single sharp pole plus higher excited states and continuum.
All the hadrons $|n\rangle$ carrying the same quantum numbers as the current
$J(x)$ will give contributions to $\text{Im}\Pi(s)$ in Eq.~\eqref{Imaginary}.
To pick out the lowest lying resonance in a particular channel, we define moments
by taking derivatives of the correlation function $\Pi(q^2)$ in Euclidean region $Q^2=-q^2>0$ \cite{1985-Reinders-p1-1,2014-Chen-p201-215}:
\begin{align}
M_n(Q^2_0)=\frac{1}{n!}\bigg(-\frac{d}{dQ^2}\bigg)^n\Pi(Q^2)|_{Q^2=Q_0^2}
=\int_{16m_Q^2}^{\infty}\frac{\rho(s)}{(s+Q^2_0)^{n+1}}ds\, ,
\label{moment}
\end{align}
where the spectral function $\rho(s)=\text{Im}\Pi(s)/\pi$.
Using Eq.~(\ref{Imaginary}), we can write the moments as
\begin{align}
M_n(Q^2_0)=\frac{f_X^2}{(m_X^2+Q_0^2)^{n+1}}\big[1+\delta_n(Q_0^2)\big],
\label{Phemoment}
\end{align}
in which $\delta_n(Q_0^2)$ contains the contributions of higher states and the continuum.
The behaviour of $\delta_n(Q_0^2)$ depends on both $Q_0^2$ and $n$. For a certain
value of $Q_0^2$, $\delta_n(Q_0^2)$ tends to zero as $n$ goes to infinity, but this convergence
becomes slower for larger $Q_0^2$.
To eliminate $f_X$ in Eq.~\eqref{Phemoment}, we consider the following ratio of the moments
\begin{align}
r(n,Q_0^2)\equiv\frac{M_{n}(Q_0^2)}{M_{n+1}(Q_0^2)}=\big(m_X^2+Q_0^2\big)
\frac{1+\delta_{n}(Q_0^2)}{1+\delta_{n+1}(Q_0^2)}. \label{ratio}
\end{align}
For sufficiently large $n$, one expects that $\delta_{n}(Q_0^2)\cong\delta_{n+1}(Q_0^2)$
for convergence \cite{1985-Reinders-p1-1}. Then we can immediately extract the hadron mass of the lowest
lying resonance $m_X$ from the above ratio $r(n,Q_0^2)$
\begin{align}
m_X=\sqrt{r(n,Q_0^2)-Q_0^2}\, . \label{mass}
\end{align}

The two-point correlation function $\Pi(q^2)$ and moments $M_n(Q^2_0)$ can also be
evaluated at the quark-gluon level using the operator product expansion (OPE) method.
For these doubly hidden-charm/bottom systems, all QCD condensates involving the
heavy quark fields do not contribute to $\Pi(q^2)$,
and we only need to calculate the perturbative term and gluon condensate in the OPE series.
The full expressions of the correlation function
$\Pi(q^2)$ and moments $M_n(Q^2_0)$ are too lengthy to be shown here.
We only list the result for the interpolating current $J_1(x)$ with $J^{PC}=0^{++}$ as an example
\begin{align}
\nonumber
\Pi^{pert}(Q^2)=&\frac{1}{128\pi^6}\int_0^1dx\int_0^1dy\int_0^1dz\Bigg\{\left[\frac{3Q^2(1-x)yz}{(1-z)^2}
+\frac{2m_Q^2}{(1-z)^3}\right]\times
\\ \nonumber & 4xy^2z(1-y)F(m_Q^2, Q^2)^3-\frac{3xy^3z(1-x)(1-y)}{(1-z)^3}F(m_Q^2, Q^2)^4
\\ \nonumber & -\left[\frac{m_Q^4y}{(1-z)^2}+\frac{Q^4xy^3z^3(1-x)(1-y)}{(1-z)}+
\frac{2m_Q^2Q^2xy^2z^2(1-y)}{(1-z)^2}\right]\times
\\ \nonumber & 6F(m_Q^2, Q^2)^2\Bigg\}\log{\left[F(m_Q^2, Q^2)\right]}\, ,
\\  \label{Pi_expression}
\Pi^{\GGa}(Q^2)=&\frac{\GGb}{512\pi^6}\int_0^1dx\int_0^1dy\int_0^1dz\Bigg\{
\\ \nonumber & \frac{2m_Q^4}{3x^2}\left[\frac{3x}{1-y}-\frac{12}{y}+8(1-x)z+\frac{8x^3(1-y)z}{(1-x)^3y}\right]
\\ \nonumber & -\frac{16m_Q^2(1-x)(1-y)z}{x^2}F(m_Q^2, 2Q^2)+\frac{3(1-x)y^2z}{1-z}F(m_Q^2, Q^2)^2
\\ \nonumber & -m_Q^2\left[\frac{2xyz^2(1-y)}{(1-x)(1-z)^2}+\frac{yz}{1-z}-\frac{8x(1-y)z}{(1-x)^2(1-z)}\right]
F(2m_Q^2, 3Q^2)
\\ \nonumber &-6Q^2(1-x)y^2z^2F(m_Q^2, Q^2)+Q^4(1-x)y^2z^3(1-z)\Bigg\}\log{\left[F(m_Q^2, Q^2)\right]}
\\ \nonumber &-\frac{m_Q^2\GGb}{192\pi^6}\int_0^1dx\int_0^1dy\int_0^1dz\Bigg\{\frac{1-z}{F(m_Q^2, Q^2)}\times
\\ \nonumber & \left[\frac{m_Q^2[Q^2(1-x)(1-y)yz^2+m_Q^2]}{x^3y^2}+\frac{Q^2(1-x)z^2[Q^2(1-y)(1-z)z+m_Q^2]}{x^2}\right] \Bigg\} \, ,
\end{align}
where $F(am_Q^2, \,bQ^2)=a\left[\frac{m_Q^2(1-z)}{1-y}+\frac{m_Q^2(1-z)}{(1-x)xy}+m_Q^2z\right]-bQ^2z(1-z)$.

To perform the numerical analysis, we use the following parameter values of the heavy
quark masses (``running'' masses in the $\overline{\rm MS}$ scheme) and the gluon condensate
\cite{2014-Olive-p90001-90001,1996-Narison-p162-172,2003-Ioffe-p229-241,2010-Narison-p559-559}
\begin{align}
\nonumber m_c(\overline{\rm MS})&=1.27\pm0.03~\text{GeV}, \,
\\
\label{parameters} m_b(\overline{\rm MS})&=4.18\pm0.03~\text{GeV}\, ,
\\
\nonumber \GGb&=(0.48\pm0.14)~\text{GeV}^4\, .
\end{align}

Using these parameter values, we can perform
the numerical analysis of the $cc\bar c\bar c$ and $bb\bar b\bar b$ systems. There are two important parameters
in the ratio $r(n,Q_0^2)$ defined in Eq.~\eqref{ratio}: $n$ and $Q_0^2$. The original literature used the
moment at $Q_0^2=0$ to perform moment method analysis \cite{1979-Shifman-p385-447,1979-Shifman-p448-518}. However, it was emphasized that the
higher dimensional condensates in the OPE series give large
contribution to the correlation function and moments at the point $Q_0^2=0$ \cite{1981-Reinders-p109-109,
1985-Reinders-p1-1}, and thus lead to bad OPE convergence and unreliable mass predictions.
In this letter, we use $Q_0^2>0$ to ensure good OPE convergence.
Thus, we need to carefully choose the two parameters $n$ and $Q_0^2$. For simplicity, we define $\xi=Q^2_0/16m_c^2$ and $Q^2_0/m_b^2$ for $cc\bar c\bar c$ and $bb\bar b\bar b$, respectively.
These two parameters, $n$ and $\xi$, are correlated with each other:
\begin{enumerate}

\item As mentioned above, a larger value of $\xi$ means slower convergence
of $\delta_{n}(Q_0^2)$. Then it will be difficult to extract the mass of the lowest lying resonance
in Eq.~\eqref{ratio}. This can be compensated by taking higher derivative $n$ of $\Pi(q^2)$ for the
lowest lying resonance to dominate.

\item However, a large value of $n$ means moving further away from the asymptotically free region. The OPE
convergence would also become bad.

\end{enumerate}

In this letter, we use the interpolating current $J_1(bb\bar b\bar b)$ with $J^{PC}=0^{++}$ as an example
to illustrate our calculations and analyses for the doubly hidden-bottom $bb\bar b\bar b$ system.
The upper limit of $n$ can be obtained by studying the convergence of the OPE series.
Requiring the perturbative term to be larger than the gluon condensate term, we obtain upper limits
$n_{max}=75, 76, 77, 78$ for  $\xi=0.2, 0.4, 0.6, 0.8$, respectively. The OPE fails for $n>n_{max}$.
The upper bound $n_{max}$ increases with respect to the value of $\xi$. In principle, one can arrive
at a region in the $(n, \xi)$ plane where the lowest lying resonance dominates the moments in
Eq.~(\ref{moment}) and the OPE series has good convergence. We will perform our analysis in such
$(n, \xi)$ parameter regions.
\begin{figure}[hbt]
\begin{center}
\scalebox{0.65}{\includegraphics{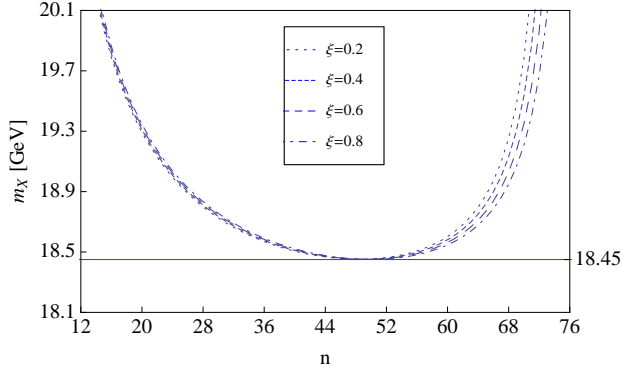}}
\caption{Hadron mass $m_{X_b}$ for $J_{1}(bb\bar b\bar b)$ with $J^{PC}=0^{++}$, as a function of $n$
for different value of $\xi$.} \label{figbbbb0++}
\end{center}
\end{figure}
\begin{figure}[hbt]
\begin{center}
\scalebox{0.55}{\includegraphics{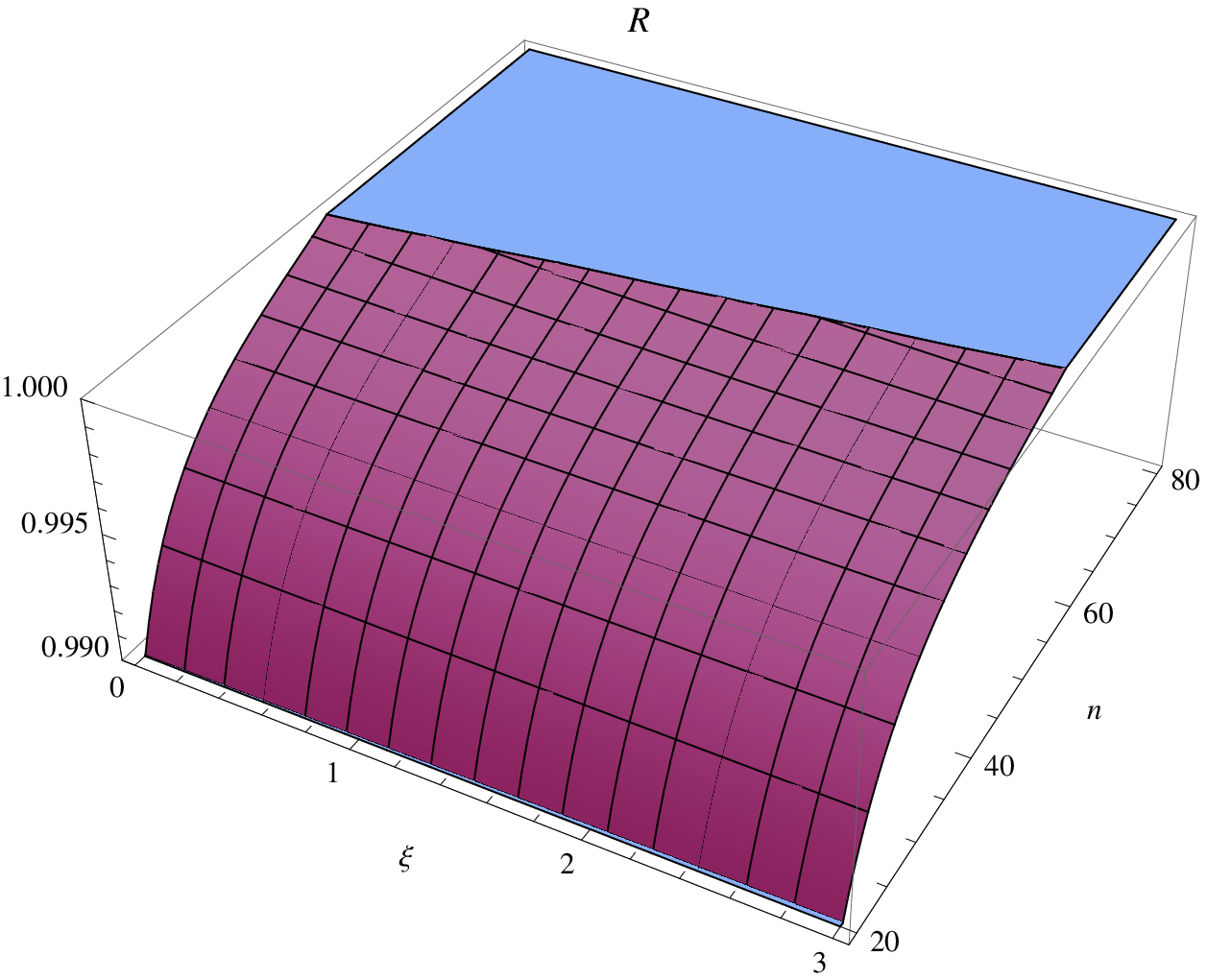}}
\caption{Ratio $R$ as a function of $n$ and $\xi$, for $J_{1}(bb\bar b\bar b)$ with $J^{PC}=0^{++}$.}
\label{fig:inequality}
\end{center}
\end{figure}
The hadron mass in Eq. \eqref{mass} is extracted as a function of $n$ and $\xi$. In Fig.~\ref{figbbbb0++},
we show the variation of the hadron mass $m_{X_b}$ with respect to $n$ for $\xi=0.2, 0.4, 0.6, 0.8$.
The mass curves have plateaus which provide stable mass predictions.
We find that these plateaus can be obtained by studying the integral expression of the moments
$M_n(Q^2_0)$ in Eq.~\eqref{moment}, which satisfies the Schwarz inequality \cite{1999-Steele-p201-206}
(as a special case of H\"older's inequality) in the following relation
\begin{align}
R=\frac{M_n(Q^2_0)^2}{M_r(Q^2_0)M_{2n-r}(Q^2_0)}\leq 1\, , \label{eq:inequality}
\end{align}
in which $r<2n$. In Fig.~\ref{fig:inequality}, we show the ratio $R$ as a function of $n$ and $\xi$,
in which $R>1$ in the blue region while $R\leq 1$ in the pink region. The demarcation line between
these two regions gives the values of $(n, \xi)$ for the plateaus in the mass curves. We thus obtain
the plateaus $(n, \xi)=(48, 0.2), (49, 0.4), (49, 0.6), (50, 0.8)$ for the mass curves in Fig.~\ref{figbbbb0++}.
One notes that the inequalities in Eq. \eqref{eq:inequality} provide much stronger constrains for
the $(n, \xi)$ plane than the OPE convergence.
Finally, we obtain the hadron mass
\begin{equation}
m_{X_b}=(18.45 \pm 0.15) \,\mbox{GeV}\, ,
\end{equation}

\begin{figure}[hbt]
\begin{center}
\scalebox{0.53}{\includegraphics{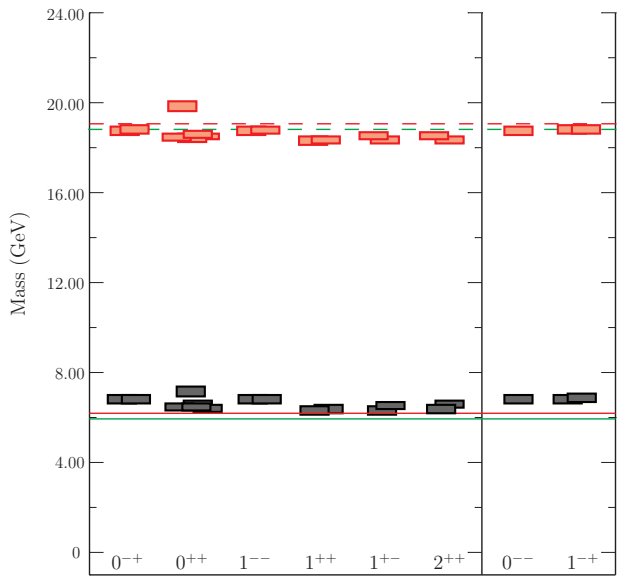}}
\end{center}
\caption{Summary of the doubly hidden-charm/bottom tetraquark spectra labelled by $J^{PC}$.
The red and black rectangles are the masses of the $cc\bar c\bar c$ and $bb\bar b\bar b$ states,
respectively. The vertical size of the rectangle represents the uncertainty of our calculation.
The green and red solid (dashed) lines indicate the two-charmonium
(bottomonium) thresholds $\eta_c(1S)\eta_c(1S)$ ($\eta_b(1S)\eta_b(1S)$) and $J/\psi J/\psi$
($\Upsilon(1S)\Upsilon(1S)$), respectively.} \label{fig:spectra}
\end{figure}

\begin{table}
\begin{center}
\begin{tabular*}{8.4cm}{cccc}
\hline
~~~~$J^{PC}$ ~~~~& Currents &~~~~ $m_{X_{c}}$\mbox{(GeV)} ~~~~&~~~~ $m_{X_{b}}$\mbox{(GeV)} ~~~~ \\
\hline
$0^{++}$      & $J_1$               &  $6.44\pm0.15$                  & $18.45\pm0.15$  \\
              & $J_2$               &  $6.59\pm0.17$                & $18.59\pm0.17$        \\
                & $J_3$               &  $6.47\pm0.16$                &$18.49\pm0.16$     \\
                    & $J_4$               &  $6.46\pm0.16$                 &$18.46\pm0.14$       \\
                       & $J_5$               &  $6.82\pm0.18$                  & $19.64\pm0.14$
\vspace{4pt}\\
$0^{-+}$        & $J_1^+$            &  $6.84\pm0.18$                 & $18.77\pm0.18$ \\
                     & $J_2^+$            &  $6.85\pm0.18$                 & $18.79\pm0.18$
\vspace{4pt}\\
$0^{--}$       & $J_1^-$              &  $6.84\pm0.18$                & $18.77\pm0.18$
\vspace{4pt}\\
$1^{++}$      & $J_{1\mu}^+$    & $6.40\pm0.19$                & $18.33\pm0.17$  \\
                    & $J_{2\mu}^+$    & $6.34\pm0.19$                & $18.32\pm0.18$
\vspace{4pt}\\
$1^{+-}$       & $J_{1\mu}^-$     & $6.37\pm0.18$                & $18.32\pm0.17$  \\
                    & $J_{2\mu}^+$    & $6.51\pm0.15$                & $18.54\pm0.15$
\vspace{4pt}\\
$1^{-+}$       & $J_{1\mu}^+$    & $6.84\pm0.18$                 & $18.80\pm0.18$ \\
                    & $J_{2\mu}^+$    & $6.88\pm0.18$               & $18.83\pm0.18$
\vspace{4pt}\\
$1^{--}$        & $J_{1\mu}^-$     & $6.84\pm0.18$                & $18.77\pm0.18$  \\
                    & $J_{2\mu}^-$     & $6.83\pm0.18$              & $18.77\pm0.16$
\vspace{4pt}\\
$2^{++}$       & $J_{1\mu\nu}$  & $6.51\pm0.15$                & $18.53\pm0.15$  \\
                     & $J_{2\mu\nu}$  & $6.37\pm0.19$                & $18.32\pm0.17$  \\
\hline
\end{tabular*}
\caption{Masses of the doubly hidden-charm $cc\bar c\bar c$ and doubly hidden-bottom $bb\bar b\bar b$
tetraquarks with various quantum numbers. \label{tablemass}}
\end{center}
\end{table}
Using the interpolating currents in Eqs.~\eqref{currents1}--\eqref{currents5}, we investigate all these
channels and collect the masses of these $cc\bar c\bar c$ and $bb\bar b\bar b$ tetraquark states in
Table \ref{tablemass}. The errors come from the uncertainties of $\xi$, the heavy quark masses and
the gluon condensate in Eq.~\eqref{parameters}.
We find that the positive parity states are lighter than the negative parity states.
Comparing these states to the two-meson mass thresholds in Fig.~\ref{fig:spectra}, one notes that
the $bb\bar b\bar b$ tetraquark states lie below the mass thresholds of $\Upsilon(1S)\Upsilon(1S)$
and $\eta_b(1S)\eta_b(1S)$, except one highest result for $0^{++}$ state. Although all $cc\bar c\bar c$ tetraquarks
lie above the two-charmonium thresholds, some of the positive parity states ($J^{PC}=0^{++}, 1^{++},
1^{+-}, 2^{++}$) are very close to the mass of the $J/\psi J/\psi$. These results for the $cc\bar c\bar c$
tetraquarks are compatible with those predicted in Ref. \cite{1975-Iwasaki-p492-492} and by a quark-gluon
model \cite{1981-Chao-p317-317}, potential model \cite{1982-Ader-p2370-2370} and the parametrized Hamiltonian approach \cite{2004-Lloyd-p14009-14009}. In the framework of diquark model, the masses of the $cc\bar c\bar c$
and $bb\bar b\bar b$ tetraquarks with $J^{PC}=0^{++}, 1^{+-}, 2^{++}$ were also obtained by solving nonrelativistic Schroedinger equation \cite{2012-Berezhnoy-p34004-34004}. However, their masses for the $cc\bar c\bar c$
states are slightly lower than our results while a bit higher for the $bb\bar b\bar b$ tetraquarks.

In general, the $QQ\bar Q\bar Q$ states mainly decay into two $Q\bar Q$ meson final states by spontaneous
dissociation, like the decay properties for the light scalar mesons \cite{2016-Rupp-p-}.
For the doubly hidden-charm $cc\bar c\bar c$ tetraquarks, they can decay via the
spontaneous dissociation mechanism since they lie above the two-charmonium thresholds.
For the doubly hidden-bottom $bb\bar b\bar b$ states, they lie below any two-bottomonium
threshold and thus can not decay into two bottomonium mesons.
Considering the restrictions of the symmetries, we collect their possible dissociation decay channels in Table \ref{ccccdecay} for
both $S$-wave and $P$-wave.
\begin{table}[hbt]
\begin{center}
\begin{tabular*}{9.2cm}{ccc}
\hline
$J^{PC}$       & S-wave              & P-wave   \\
\hline
$0^{++}$  & $\eta_c(1S)\eta_c(1S)$, $ J/\psi J/\psi$ &  $\eta_c(1S)\chi_{c1}(1P)$, $ J/\psi h_c(1P)$
\vspace{6pt}\\
$0^{-+}$        & $\eta_c(1S)\chi_{c0}(1P)$, $ J/\psi  h_c(1P)$    &  $ J/\psi J/\psi$
\vspace{6pt}\\
$0^{--}$         & $ J/\psi\chi_{c1}(1P)$    &  $ J/\psi\eta_c(1S)$
\vspace{6pt}\\
$1^{++}$       &  $-$   &  $ J/\psi h_c(1P)$,  $\eta_c(1S)\chi_{c1}(1P)$,  \\
                     &                                                 &  $\eta_c(1S)\chi_{c0}(1P)$
\vspace{6pt}\\
$1^{+-}$        & $ J/\psi\eta_c(1S)$      &  $ J/\psi\chi_{c0}(1P)$, $ J/\psi\chi_{c1}(1P)$,  \\
                     &                                                &  $\eta_c(1S) h_c(1P)$
\vspace{6pt}\\
$1^{-+}$  & $ J/\psi h_c(1P)$, $\eta_c(1S)\chi_{c1}(1P)$ &   $-$
\vspace{6pt}\\
$1^{--}$         & $ J/\psi\chi_{c0}(1P)$, $ J/\psi\chi_{c1}(1P)$,     &  $ J/\psi\eta_c(1S)$ \\
                     & $\eta_c(1S) h_c(1P)$                                                                &
\\ \hline
\end{tabular*}
\caption{Possible decay modes of the $cc\bar c\bar c$ states by spontaneous dissociation
into two charmonium mesons. \label{ccccdecay}}
\end{center}
\end{table}

These $QQ\bar Q\bar Q$ tetraquarks can also decay into a doubly charmed/bottomed baryon ($QQq$) pair by
the creation of a light quark pair $QQ\bar Q\bar Q\to (QQq) + (\bar Q\bar Q\bar q)$ so long as the kinematics allows.
However, the thresholds of two doubly charmed/bottomed baryons
are higher than the extracted masses of the $cc\bar c\bar c$ and $bb\bar b\bar b$ states in Table \ref{tablemass} \cite{2002-Mattson-p112001-112001}.
Hence, such decays are forbidden by the kinematics.
The two singly heavy meson pair decay modes $QQ\bar Q\bar Q\to (q\bar Q) + (Q\bar q)$ are possible
via a heavy quark pair annihilation and a light quark pair creation at the same time. In general, these decays
should be suppressed by the annihilation of a heavy quark pair. However, such suppression can be compensated
by the large phase space factor \cite{1981-Chao-p317-317}. Consequently, the two heavy meson decay modes
will contribute significantly to the total width of the doubly hidden-charm/bottom tetraquark states.

The doubly hidden-charm and hidden-bottom tetraquark states lie in the energy region much higher than
the conventional charmonium and bottomonium mesons, respectively. They can be clearly distinguished
experimentally from the normal $q\bar q$ states. The production of the $QQ\bar Q\bar Q$ states is extremely
difficult because that requires two heavy quark pairs be created. However, the recent observations of the
$J/\psi$ pair~\cite{2012-Aaij-p52-59,2014-Khachatryan-p94-94} and the simultaneous
$J/\psi\Upsilon(1S)$~\cite{2016-Abazov-p82002-82002,CMS1} and $\Upsilon(1S)\Upsilon(1S)$
events~\cite{2017-Khachatryan-p13-13} shed some light for the production of these doubly
hidden-charm/bottom tetraquarks.
Thus, the $J/\psi J/\psi$ and $\eta_c(1S)\eta_c(1S)$ channels can be good choices to search for the doubly
hidden-charm $cc\bar c\bar c$ states, and LHCb, D0, and CMS are potential and ideal platforms.

To end this paper, we suggest to search for the doubly hidden-charm states in the $J/\psi J/\psi$ and $\eta_c(1S)\eta_c(1S)$ channels.
Our predictions
will provide new possible phenomena in hadron colliders. In the near future, we hope that these doubly
hidden-charm/bottom tetraquark states can be observed at facilities such as LHCb, D0, CMS, RHIC and the
forthcoming BelleII, where many heavy quarks are produced.

This project is supported by the Natural Sciences and Engineering Research Council of
Canada (NSERC) and the National Natural Science Foundation of China under Grants 11205011,
No. 11475015, No. 11375024, No. 11222547, No. 11175073, and No. 11575008; the Ministry of
Education of China (SRFDP under Grant No. 20120211110002 and the Fundamental Research
Funds for the Central Universities). X.L. is also supported by the National Youth Top-notch
Talent Support Program (Thousands-of-Talents Scheme).


\end{document}